\documentclass[%
 reprint,
 amsmath,amssymb,
 aps,
]{revtex4-2}

\usepackage{graphicx}% Include figure files
\usepackage{dcolumn}% Align table columns on decimal point
\usepackage{bm}% bold math

\newcommand{\bea}{\begin{eqnarray}}
\newcommand{\eea}{\end{eqnarray}}
\newcommand{\be}{\begin{equation}}
\newcommand{\ee}{\end{equation}}

\begin{document}

%\preprint{APS/123-QED} 

\title{The connection between regular black holes in nonlinear electrodynamics and semi-classical dust collapse}

\author{Daniele Malafarina}
\email{daniele.malafarina@nu.edu.kz}
\affiliation{Department of Physics, Nazarbayev University, Kabanbay Batyr 53, 010000 Nur-Sultan, Kazakhstan.}%

\author{Bobir Toshmatov}
\email{toshmatov@astrin.uz}
\affiliation{Ulugh Beg Astronomical Institute, Astronomy str. 33, Tashkent 100052, Uzbekistan}
%\affiliation{Akfa University, National Park str., Tashkent 100095, Uzbekistan}
\affiliation{Institute of Nuclear Physics, Ulugbek 1, Tashkent 100214, Uzbekistan}

\date{\today}% It is always \today, today,
             %  but any date may be explicitly specified

\begin{abstract}
There exist a correspondence between black holes in non linear electrodynamics (NLED) and gravitational collapse of homogeneous dust with semi-classical corrections in the strong curvature regime that to our knowledge has not been noticed until now. We discuss the nature of such correspondence and explore what insights may be gained from considering black holes in NLED in the context of semi-classical dust collapse and vice-versa. 
\end{abstract}

%\keywords{Suggested keywords}%Use showkeys class option if keyword
                              %display desired
\maketitle

%\section{Article}\label{sec1}

{\em Static black holes and dust collapse:}\\
It is well known that a radially infalling particle in the Schwarzschild spacetime follows the same trajectory as a particle in the gravitational collapse of a non-rotating homogeneous dust sphere \cite{OSD}. This is already true in Newtonian gravity in fact and the best way to illustrate it is to express the Schwarzschild line element in Lemaitre coordinates $\{\rho,\tau\}$ \cite{Lemaitre}. These are the coordinates measured by an observer in free fall.
For a static and spherically symmetric space-time given in Schwarzschild coordinates $\{R,T\}$ by \be \label{metric}
ds^2=-f(R)dT^2+\frac{dR^2}{f(R)}+R^2d\Omega^2 \, ,
\ee 
they are obtained from the transformation $R=R(\rho,\tau)$, $T=T(\rho,\tau)$ given by
\bea 
d\tau&=& dT+\frac{g(R)}{f(R)}dR \, , \\
d\rho&=&dT+\frac{1}{g(R)f(R)}dR \, ,
\eea 
with $g=\sqrt{1-f}$, so that
if we take
\be 
f(R)=1-\frac{2\mathcal{M}(R)}{R} \, ,
\ee 
we get
\be \label{metric1}
ds^2=-d\tau^2+\frac{2\mathcal{M}(R)}{R}d\rho^2+R(\tau,\rho)^2d\Omega^2 \, .
\ee

For Schwarzschild we have $\mathcal{M}(R)=M={\rm const}.$ and a particle in free fall at $\rho=\rho_0$ follows the trajectory $B_0(\tau)=R(\rho_0,\tau)$. From the change of coordinates we have
\be 
d\rho-d\tau=\frac{1}{g}dR=\sqrt{\frac{R}{2\mathcal{M}(R)}}dR=\sqrt{\frac{R}{2M}}dR \, ,
\ee 
which for $\rho=\rho_0$ implies $d\rho_0=0$ and gives %the following equation for $B_0$
\be \label{0}
\frac{dB_0}{d\tau}=-\sqrt{\frac{2\mathcal{M}(B_0)}{B_0}}=-\sqrt{\frac{2M}{B_0}} \, ,
\ee
that once integrated with the initial condition $B_0(0)=\rho_0$ gives
\be \label{lem}
B_0(\tau)=\rho_0\left(1-\frac{3}{2}\sqrt{\frac{2M}{\rho_0^3}}\tau\right)^{2/3}=\rho_0a(\tau) \, .
\ee 
Eq.~\eqref{lem} is formally identical to the equation of motion for marginally bound homogeneous dust collapse. 

{\em NLED black holes in Lemaitre coordinates:}\\
Non-singular extensions of black hole spacetimes have been widely considered in the context of modifications to General Relativity (GR), most notably Loop Quantum Gravity (LQG) \cite{qg-bh} and NLED\cite{Beato}. Here we focus on GR coupled to NLED which can be described by the action
\be 
\mathcal{A}=\frac{1}{16\pi}\int d^4x\sqrt{|g|}\left(\mathrm{R}-\mathcal{L}_{\rm NLED}(\mathrm{F})\right) \, ,
\ee 
with the Lagrangian given by
\be 
\mathcal{L}_{\rm NLED}=\frac{4\mu}{\alpha}\frac{(\alpha\mathrm{F})^{(\nu+3)/4}}{[1-(\alpha \mathrm{F})^{\nu/4}]^{1+\mu/\nu}} \, ,
\ee 
where $\alpha$ is the coupling to NLED and $\mathrm{F}=\mathrm{F}_{\kappa\lambda}\mathrm{F}^{\kappa\lambda}$ is the Faraday tensor of the electromagnetic field \cite{NLED}. Since there is no matter Lagrangian the energy momentum tensor in Schwarzschild coordinates is due solely to $\mathcal{L}_{\rm NLED}$ as
\be 
T_{\kappa\lambda}=\frac{1}{4\pi}\left(\partial_{\mathrm{F}}\mathcal{L}_{\rm NLED}\mathrm{F}^\sigma_{\kappa}\mathrm{F}_{\lambda\sigma}-\frac{1}{4}g_{\kappa\lambda}\mathcal{L}_{\rm NLED}\right) \, ,
\ee 
which gives
\bea 
T_0^0&=&T_1^1=-\frac{2 \mathcal{M}'(R)}{R^2}\, , \\
T_2^2&=&T_3^3=-\frac{\mathcal{M}''(R)}{R}\, .
\eea 
For a static and spherically symmetric spacetime with line element Eq.~\eqref{metric} and NLED source given by a magnetic charge $q_*$ we obtain a black hole solution with
\be \label{f}
\mathcal{M}(R)=\frac{MR^{\mu}}{(R^\nu+q^\nu_*)^{\mu/\nu}} \, .
\ee 
Notice that for $\mu=0$ we retrieve the Schwarzschild solution while we need to impose $\mu\geq 3$ in order for the solution to be regular at $R=0$ \cite{Fan}.

We can then follow the same procedure outlined for Schwarzschild, and moving to Lemaitre coordinates we get
\be 
g(R)=\sqrt{\frac{2MR^{\mu-1}}{(R^\nu+q_*^\nu)^{\mu/\nu}}} \, .
\ee
The equation of motion for a free falling observer at $\rho=\rho_0={\rm const.}$ becomes
\be \label{1}
\frac{dB_0}{d\tau}=-\sqrt{\frac{2MB_0^{\mu-1}}{(B_0^\nu+q_*^\nu)^{\mu/\nu}}}=-\sqrt{\frac{2M}{B_0}\left(1+\frac{q_*^\nu}{B_0^\nu}\right)^{-\mu/\nu}} \, .
\ee 
If we consider the scaling $B_0(\tau)=\rho_0a(\tau)$ and define $2M=m_0\rho_0^3$ and $q_*=\rho_0q$ the above equation becomes
\be \label{2}
\frac{da}{d\tau}=-\sqrt{\frac{m_0}{a}\left(1+\frac{q^\nu}{a^\nu}\right)^{-\mu/\nu}} \, ,
\ee 
which looks very similar to the equation of motion of semi-classical dust collapse models.

{\em Dust collapse with semi-classical corrections:}\\
Marginally bound homogeneous dust collapse, also known as the Oppenheimer-Snyder-Datt (OSD) model \cite{OSD}, is obtained by solving the field equations for an homogeneous fluid sphere with vanishing pressures. Semi-classical corrections to collapse models have been considered mostly in the context of scalar fields \cite{scalar-field} but the formalism is readily extended to dust and homogeneous perfect fluids and strong field corrections have been developed in a variety of different settings \cite{review}.
The action for the semi-classical dust collapse is
\be 
\mathcal{A}=\frac{1}{16\pi}\int d^4x\sqrt{|g|}\left(\mathrm{R}-\mathcal{L}_{\rm Dust}-\mathcal{L}_{\rm corr}\right) \, ,
\ee 
where $\mathcal{L}_{\rm corr}$ is the Lagrangian density of the strong curvature corrections to the theory, from which we obtain the effective energy momentum tensor
\be 
\mathrm{T}_{\kappa\lambda}^{\rm eff}=\mathrm{T}_{\kappa\lambda}^{\rm Dust}+\mathrm{T}_{\kappa\lambda}^{\rm corr} \, ,
\ee 
where the dust part is given simply by $\mathrm{T}^{\kappa\lambda}_{\rm Dust}=\epsilon u^\kappa u^\lambda$, with $u^\kappa$ being the 4-velocity of the fluid, while $\mathrm{T}_{\kappa\lambda}^{\rm corr}$ describes the strong field corrections as an unphysical addition to the energy-momentum tensor. In general we may expand $\mathrm{T}_{\kappa\lambda}^{\rm corr}$ in powers of $\epsilon$ close to $\epsilon\simeq 0$ (i.e for $\epsilon$ small with respect to some critical density $\epsilon_{\rm cr}$) and write $\mathrm{T}_{\kappa\lambda}^{\rm eff}$ as
\be \label{eps-eff}
\epsilon_{\rm eff}=\epsilon+\alpha_1\epsilon^2+\alpha_2\epsilon^3+... \, .
\ee 
Notice that the effective energy-momentum tensor is not dust anymore since, while $p=0$, we now have $p_{\rm corr}\neq 0$. In fact it is the effective pressure that allows for violation of the energy conditions and consequently may halt collapse before it reaches the singularity. 

The line element for spherically symmetric, marginally bound, homogeneous collapse is simply
\be 
ds^2=-dt^2+B'^2dr^2+B^2d\Omega^2 \, ,
\ee 
with $B=B(r,t)$ and primed quantities representing partial derivatives with respect to $r$, i.e. $X'=\partial X /\partial r$. 
In the classical case $\mathrm{T}_{\kappa\lambda}^{\rm corr}=0$ and the energy density $\epsilon$ of the collapsing sphere is given by
\be \label{epsilon}
\epsilon=\frac{F'}{B^2B'} \, ,
\ee 
with $F$ being the mass contained within a given radius $r$ at a time $t$, namely the Misner-Sharp mass, which is defined as
$F(r,t)=B\dot{B}^2$,
and dotted quantities indicating partial derivatives with respect to $t$, i.e. $\dot{X}=\partial X /\partial t$. For homogeneous dust the Misner-Sharp mass does not depend on the proper time $t$ as there is no inflow or outflow of particles through any constant $r$ surface during collapse. If we take the boundary of the cloud $r=r_0={\rm const.}$ we can see that the 3-metric restricted to the boundary surface is identical to the black hole metric in Lemaitre coordinates \eqref{metric1} with the identification of $\tau=t$ and $B(r_0,t)=B_0(t)=B_0(\tau)$ and $F(r_0)=2M$. The equation of motion for marginally bound collapse becomes
\be \label{Bdot}
\dot{B}=-\sqrt{\frac{F}{B}} \, ,
\ee 
which, when evaluated at the boundary, reduces to Eq.~\eqref{0}. It is customary to express the above equations in terms of the adimensional scale factor $a$ and rescale the Misner-Sharp mass as
\be 
B(r,t)=r a(t) \, , \;\; F(r,t)=r^3m(r,t) \, ,
\ee
so that for homogeneous dust we have $m(r)=m_0={\rm const}.$ and Eqs.~\eqref{epsilon} and \eqref{Bdot} become
\bea
\epsilon&=&\frac{3m_0}{a^3} \, ,\\ \label{adot-osd}
\dot{a}&=&-\sqrt{\frac{m_0}{a}} \, ,
\eea
and the solution of Eq.~\eqref{adot-osd} is given by $a(t)$ as in Eq.~\eqref{lem}. In order to avoid the formation of the singularity at the end of collapse a quantum-inspired model based on LQG was proposed in \cite{BMM}. The idea, mediated from Loop Quantum Cosmology \cite{lqc}, is to add an effective correction to the energy momentum tensor which describes the departure of the Quantum-Gravity theory from classical GR in the strong field, i.e. it becomes important at high densities. A general form of such a correction consistent with Eq.~\eqref{eps-eff} is
\be 
\epsilon_{\rm eff}=\epsilon\left[1-\left(\frac{\epsilon}{\epsilon_{\rm cr}}\right)^\beta\right]^\gamma=\frac{3m_{\rm eff}}{a^3} \, ,
\ee
where $\epsilon_{\rm cr}$ is a critical density scale and $m_{\rm eff}$ is the rescaled effective Misner-Sharp mass for which
\be 
m_{\rm eff}=a\dot{a}^2 \, .
\ee
The model proposed in \cite{BMM} has $\beta=\gamma=1$ but in principle other values of $\beta$ and $\gamma$ can be considered. From the above equations we can write the equation of motion as
\be \label{3}
\dot{a}=-\sqrt{\frac{m_0}{a}\left(1-\frac{a_{\rm cr}^{\nu}}{a^{\nu}}\right)^\gamma} \, ,
\ee
where we have introduced the critical scale factor $a_{\rm cr}$ from $\epsilon_{\rm cr}=3m_0/a_{\rm cr}^3$ and set $\nu=3\beta$. It is immediately clear that Eq.~\eqref{2} and Eq.~\eqref{3} are identical with the exception of the sign in front of $a_{\rm cr}$. For the case of NLED the sign of the charge can be positive or negative, while for collapse $a_{\rm cr}>0$ because it related to a length scale.

\begin{figure}[hhhh]
\includegraphics[width=0.75\columnwidth]{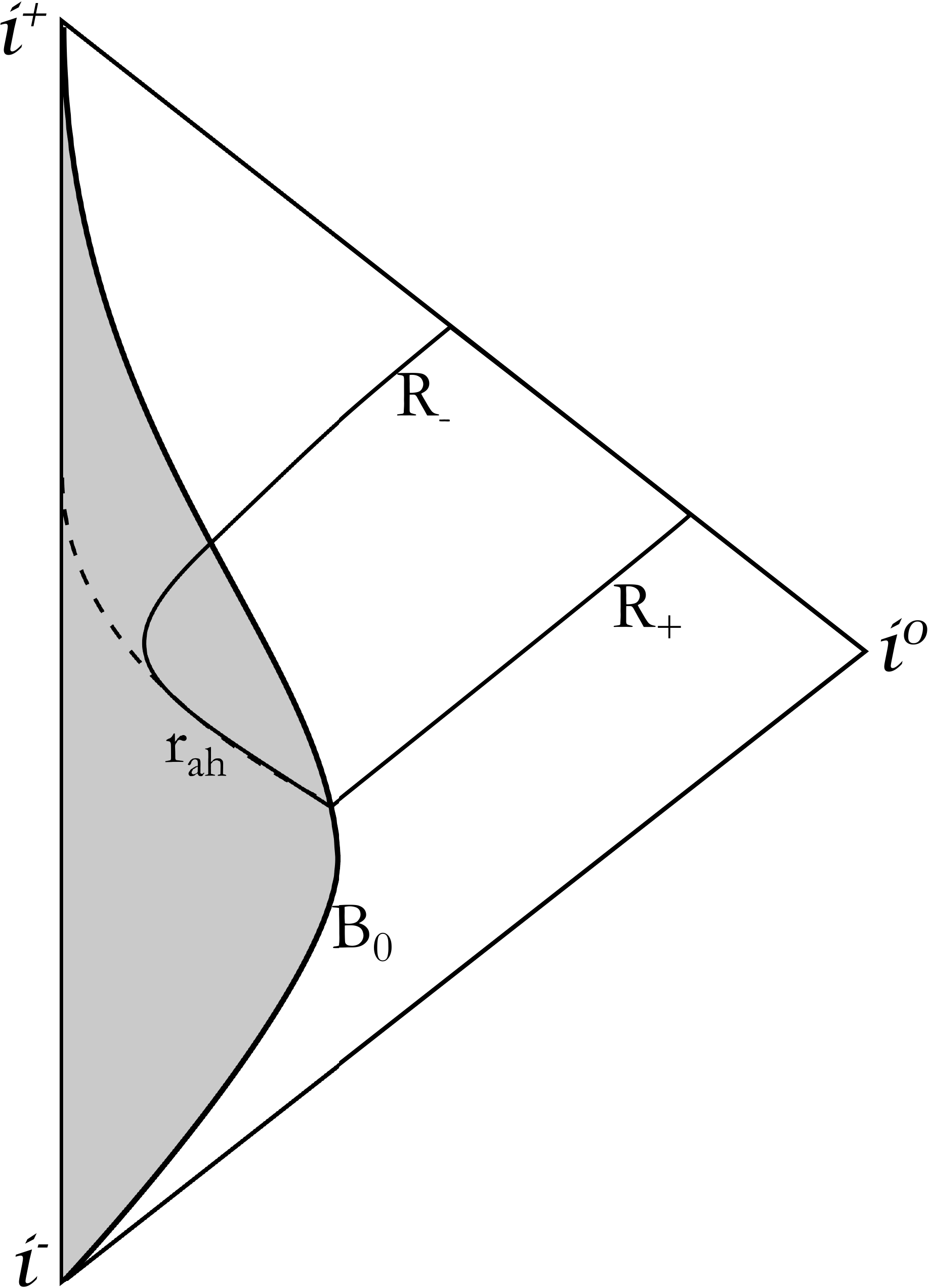}
\caption{The Penrose diagram of gravitational collapse of dust with semi-classical corrections that leads to the formation of the Hayward regular black hole. The boundary of the cloud $B_0$ collapses towards $r=0$. The apparent horizon $r_{\rm ah}$ crosses the boundary twice producing the outer and inner horizons of the Hayward solutions ($R_+$ and $R_-$ respectively). The dashed line represents the apparent horizon for the OSD collapse. }
\label{fig1}
\end{figure}

The condition for the formation of trapped surfaces is
\be 
1-\frac{F}{B}=1-\frac{r^2m_{\rm eff}}{a}=1-r^2\dot{a}^2=0 \, ,
\ee 
which gives implicitly the radius of the apparent horizon $r_{\rm ah}(t)=-1/\dot{a}$. Notice that $r_{\rm ah}(t)$ is defined only within the matter cloud and therefore the apparent horizon exists only when  $r_{\rm ah}(t)\leq r_0$. When  $r_{\rm ah}(t)=r_0$ the apparent horizon `touches' the cloud's boundary and must join with the inner or outer horizons of the exterior geometry. Assuming that there are no trapped surfaces at the initial time means that  $r_{\rm ah}(0)>r_0$ and the horizon forms only at a later stage of collapse. If $\dot{a}\rightarrow -\infty$ then $r_{\rm ah}\rightarrow 0$ and  $r_{\rm ah}(t)$ will cross the boundary only once. Therefore we can not have the formation of an inner horizon at the end of collapse. This is the case of the OSD collapse. On the other hand if $\dot{a}\rightarrow 0$ then $r_{\rm ah}$ may cross the boundary twice thus producing the outer and inner horizons (see Figure \ref{fig1}). This is the case of the Hayward regular black hole.

{\em Example 1: Hayward black hole:}\\
We can consider the Hayward black hole \cite{Hayward} by setting $\mu=\nu=3$ in Eq.~\eqref{f}, which corresponds to $\gamma=-1$. Then we may look at the corresponding semi-classical dust collapse. The equation of motion for the scale factor becomes
\be 
\dot{a}=-\sqrt{m_0\frac{a^2}{a^3+q^3}}\, .
\ee 
In this case $a\rightarrow 0$ asymptotically and the effective density remains finite as $\epsilon_{\rm eff}\rightarrow 3m_0/q^3$. The effective energy momentum tensor can be obtained from
\be 
\epsilon_{\rm eff}=\epsilon\left(1-\frac{\epsilon}{\epsilon_{\rm cr}+\epsilon}\right)\, .
\ee 
Also since $\dot{a}\rightarrow 0$ we see that, if trapped surfaces develop, the apparent horizon must cross the boundary twice, thus forming the outer and inner horizons. Finally $\ddot{a}\rightarrow 0$ shows that the collapse does not bounce and the collapsing matter asymptotically settles to the Hayward black hole. Similarly collapse to the Bardeen black hole \cite{Bardeen} can be obtained for $\mu=3$ and $\nu=2$ and other NLED black holes, such as the ones described in \cite{Beato2} may also be recast in the context of collapse.

\begin{figure}[hhh]
\centering
\includegraphics[width=0.48\textwidth]{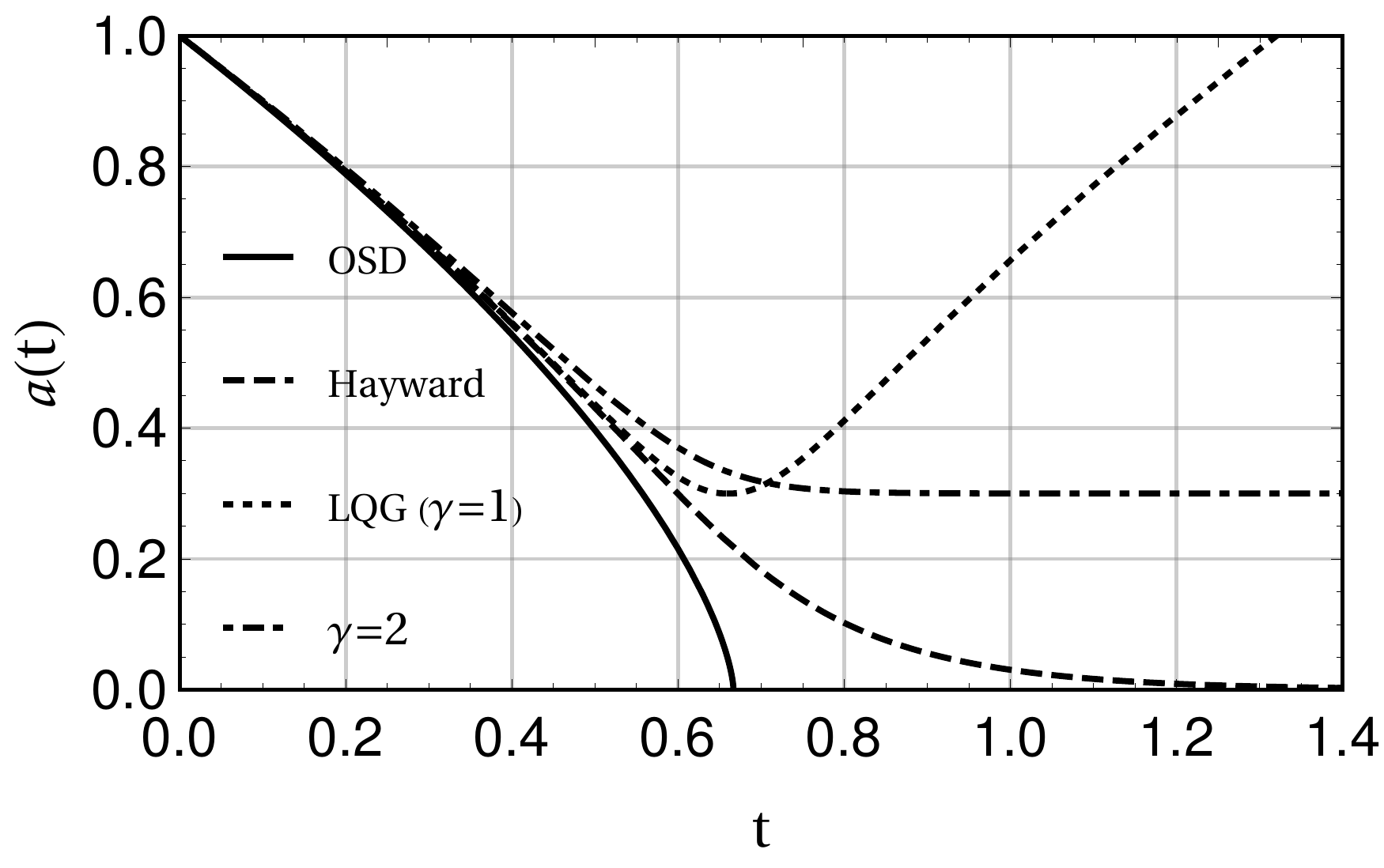}
\caption{\label{fig-at} The scale factor $a(t)$ for the OSD model and the three collapse models in the examples. The solid line describes OSD collapse. The dashed line describe collapse to a Hayward regular black hole. The dotted line describes collapse and bounce of the LQG inspired model. The dot-dashed line describes the collapse model with $\gamma=2$. In the figure we have taken $m_0=1$ and $q=-a_{\rm cr}=0.3$.}
\end{figure}

{\em Example 2: LQG-inspired collapse:}\\
Conversely we can take a model for collapse and bounce such as the one inspired by LQG from \cite{BMM} and investigate its corresponding black hole. The LQG inspired collapse is obtained from Eq.~\eqref{3} with $\beta=\gamma=1$, which correspond to $\nu=3$ and $\mu=-3$. Then the NLED Lagrangian for this black hole is
\be 
\mathcal{L}_{\rm NLED}(\mathrm{F})=-12\sqrt{\alpha}\mathrm{F}^{3/2} \, ,
\ee 
and $f(R)$ can be obtained from Eq.~\eqref{f} with $q_*=-a_{\rm cr}$.
The black hole is singular and the Kretschmann scalar diverges for $R\rightarrow 0$.
\be 
\mathcal{K}=\frac{48M^2}{R^{12}}(39q_*^6-10q_*^3R^3+R^6) \, .
\ee 
However, the equation for the radial infall of a particle in this geometry shows that it must bounce at $R=q_*$ and thus can not reach the center.

{\em Example 3: Semi-classical collapse with $\gamma=2$:}\\
From the above consideration we can easily construct a model that settles at $a_{\rm cr}$ asymptotically. If we take $\beta=1$ and $\gamma=2$ we see that while $a\rightarrow a_{\rm cr}$ we have that both $\dot{a}\rightarrow 0$ and $\ddot{a}\rightarrow 0$.
Also in this case the corresponding black hole is singular and $\mathcal{K}$ diverges at the center, but any radially infalling particle would approach $R = q_*$ in an infinite comoving time and thus this can be seen as the exterior geometry of an extreme compact object of finite size. The Kretschmann scalar for this black hole is
\be 
\mathcal{K}=\frac{48M^2}{R^{18}}(278q_*^{12}-412q_*^9R^3+180q_*^6R^6-20q_*^3R^9+R^{12}) \, ,
\ee 
while the Lagrangian density is
\be 
\mathcal{L}_{\rm NLED}(\mathrm{F})=-24\sqrt{\alpha}\mathrm{F}^{3/2}[1-(\alpha\mathrm{F})^{3/4}] \, .
\ee 

The scale factors for OSD collapse, the Hayward black hole, the LQG inspired collapse and the collapse model with $\gamma=2$ are shown in Fig.~\ref{fig-at}.

\begin{acknowledgments}
DM and BT acknowledge support from Nazarbayev University Faculty Development Competitive Research Grant No. 11022021FD2926 and Ministry of Innovative Development of the Republic of Uzbekistan Grant No.~F-FA-2021-510, respectively. 
\end{acknowledgments}

\end{document}